\documentclass[12pt]{iopart}
\usepackage{graphicx}
\usepackage{dcolumn}
\usepackage{bm}
\usepackage{braket} 
\usepackage[caption=false]{subfig} 
\usepackage{changepage}
\usepackage{float}
\usepackage{amssymb}
\usepackage{bbm}

\expandafter\let\csname equation*\endcsname\relax 
\expandafter\let\csname endequation*\endcsname\relax 
\usepackage{amsmath}

\usepackage{caption} 
\usepackage[normalem]{ulem}
\usepackage{color}
\definecolor{gray}{rgb}{.6,.6,.6}

\newcommand{\ketbra}[2]{\ensuremath{\vert\!\!\;#1\rangle \! \langle #2\vert}}
\newcommand{\qed}{\rule{1ex}{1ex}}

\begin{document}

\title{Symmetry-protection of multiphoton states of light}

\author{Jon Lasa-Alonso$^{1,2}$, Mart\'in Molezuelas$^{1}$, J. J. Miguel Varga$^{1,2}$, Aitzol Garc\'ia-Etxarri$^{2,3}$, G\'eza Giedke$^{2,3}$ and Gabriel Molina-Terriza$^{1,2,3}$}
\vspace{15pt}
\address{$^1$Centro de F\'isica de Materiales, Paseo Manuel de Lardizabal 5, 20018 Donostia-San Sebasti\'an, Spain.}
\vspace{10pt}
\address{$^2$Donostia International Physics Center, Paseo Manuel de Lardizabal 4, 20018 Donostia-San Sebasti\'an, Spain.}
\vspace{10pt}
\address{$^3$IKERBASQUE, Basque Foundation for Science, Mar\'ia D\'iaz de Haro 3, 48013 Bilbao, Spain.}
\vspace{10pt}
\ead{jlasa022@ikasle.ehu.eus; gabriel.molina.terriza@gmail.com}
\vspace{10pt}
\vspace{10pt}
\begin{indented}
\item[]May 2020
\end{indented}

\begin{abstract}
In this manuscript we analyze the emergence of protected multiphoton states in scattering problems with cylindrical symmetry. In order to do that, we first provide a formal definition of the concept of postselected symmetry-protection. We show that symmetry-protected states are not limited to one- or two-photon states, on the contrary, it can be formally extended to the multiphoton case. In addition, we prove for the case of cylindrical symmetry that all possible multiphoton protected states are constructed from a small set of one- and two-photon states. Finally, we point out possible applications that symmetry-protected states may have in quantum communications, concretely, in the construction of decoherence-free subspaces.
\end{abstract}

\section{\label{sec:level1}Introduction}

The processing of quantum information carried by photons has reached such a level of maturity that photonic quantum computers are becoming competitive in this technological field~\cite{Kok16,FSS19}. As was shown in 2001 in a seminal work \cite{KLM01}, passive linear optics, i.e. an interferometer, is sufficient for universal photonic quantum computing if combined with single-photon state preparation and feedback based on photon number measurements. More recently it was shown that, even without feedforward, these photonic devices can efficiently perform computational tasks that are supposed to be computationally hard on classical computers (``boson sampling'') \cite{aaronson2011computational, th_comp, npjQinf_rev}, something which has been demonstrated in proof-of-principle experiments \cite{WienBS, spring2013boson, spagnolo2014experimental}.

In fact, the quantum interference of photons is at the heart of the enhancement associated to quantum applications such as the processing and transmission of quantum information, which is essential to establish a quantum network of communications~\cite{Hu2016}. Quantum information can be encoded in photons within different degrees of freedom, such as transverse momentum, spatial path or time-bin, among others. In particular, the framework based on total angular momentum and helicity~\cite{helicity, PhysRevA.86.042103} has gained especial relevance due to applications such as the generation of states in high-dimensional Hilbert spaces~\cite{PhysRevLett.88.013601, fickler2016}, light-matter interactions~\cite{Zambrana-Puyalto2018}, data transmission~\cite{Wang2012}, and sensing of chirality in molecules~\cite{poulikakos2019optical, feis2020helicity, lasa2020surface}.

One fascinating feature of this framework is that it allows to describe on the same footing the paraxial and non-paraxial regimes of light~\cite{PhysRevA.86.042103,Tischler2014}. This is interesting, because most of the control of light for quantum optics experiments is performed in the paraxial regime, while light-matter interactions typically occur in subwavelength structures, such as atoms, molecules, or nanostructures. Therefore, in order to maximize the interaction in scattering problems, light beams must be strongly focused onto the samples, often leaving the paraxial regime. In fact, the study of the interaction between light and subwavelength structures is receiving a growing interest within the community~\cite{PhysRevLett.105.216802, Kuznetsovaag2472, Cheben2018, Rahmani2017, Lodahl2017}. Although the interaction of light with these structures can be described from the scattering of the electromagnetic modes, at least in the linear regime, the scattering properties of multiphoton states can be rather complex. This is due to quantum interference effects and the fact that one can equally describe multiphoton states with different sets of orthogonal modes~\cite{Silberhorn}.

In this work, we analyze the emergence of a very specific set of multiphoton states in generic scattering problems. While it is always possible to find eigenstates of a given system, i.e. states which are left invariant in the interaction with the system, these eigenstates normally depend on the particularities of the system. However, there are situations when certain states are left invariant by \emph{all} the scattering matrices compatible with certain symmetry operations. These so-called ``symmetry-protected states'' \cite{buse2018symmetry} can be non-trivial and in some situations hard to find. Here, we consider initial states of a known number of photons in a given set of angular momentum light modes and investigate their scattering on cylindrically symmetric structures. We restrict ourselves to the cases where the final state is postselected to contain all input photons in a certain set of output modes. We observe that the symmetries of the physical problem strongly constrain the possible output states. In particular, if a state is left invariant by \emph{all} the scattering matrices symmetric under rotations and mirror operations, we say that the input state is symmetry-protected in the scattering process. We also show that states that are protected in postselected scattering at cylindrically symmetric structures (Fig.~\ref{mirrors}) can only be constructed in the subspace of input states with total angular momentum equal to zero, agreeing with previous results shown in Ref.~\cite{buse2018symmetry}.

These symmetry-protected states can be useful for sensing the geometrical asymmetries present in nanostructures. Furthermore, studying these states may also pave the way to efficient transmission channels of entangled multiphoton states and decoherence-free subspaces. Actually, due to the generality of the arguments used in this work, these considerations may apply to macroscopic structures such as optical fibers, but also to nanostructures such as nanofibers~\cite{Eznaveh:18}, nanoholes~\cite{Zambrana-Puyalto2014} or nanospheres~\cite{Alabastri2016, Hirsch13549}.

The rest of the manuscript is organized as follows. After setting the general stage on the notion of symmetry-protection in Section~\ref{sec:general}, we specialize in Section~\ref{sec:ourstates} on the case of cylindrically symmetric systems and introduce the set of modes that we are going to use in this work. In Section~\ref{sec:2-photon-states} we present the results found for two-photon states, both for modes with null angular momentum and with arbitrary non-zero integer value. In Section~\ref{sec:N-photon-states} we generalize the results to an arbitrary number of photons, $N$. In Section~\ref{sec:DFS} we discuss the applications that symmetry-protected states may have in quantum communications. Finally, in Section~\ref{sec:conclusion} we summarize the main conclusions of the manuscript. 

\section{Symmetry-protection: general considerations}\label{sec:general}

We consider the scattering of a system of photons with mode space ${\mathcal H}$ on a linear passive sample that is invariant under a set of symmetry operations $G$. We denote by $T$ the full single-particle scattering matrix (usually unitary, though it may include linear losses such that $\rho\mapsto T\rho T^\dag$ is a trace-nonincreasing completely positive map) and its Fock space representation by $\hat{T}$. For a subspace ${\mathcal H}_s\subset {\mathcal H}$ of modes we denote by ${\mathcal H}_s^N$ the space of $N$ photons in the modes ${\mathcal H}_s$ and the isometry from the full Fock space to ${\mathcal H}_s$ by
\begin{equation}
\nonumber
P_s^N =
\sum_{n_1,...,n_{M-1}=0}^N \ketbra{n_1,\dots,n_{M-1},n_M}{n_1,\dots,n_{M-1},n_M},
\end{equation}
with $n_M=N-n_1-n_2-...-n_{M-1}$. The $(N,{\mathcal H}_s)$-postselected scattering matrix is defined as:
\begin{equation}
\nonumber
\hat{S} \equiv P_s^N\hat{T}(P_s^N)^\dag,
\end{equation}
which describes the quantum operation acting on ${\mathcal H}_s^N$ obtained after scattering, conditioned on finding all $N$ photons again in the modes in ${\mathcal H}_s$.

We call a $N$-photon state, $\ket{\psi} \in {\mathcal H}_s^N$, (${\mathcal H}_s$)-symmetry-protected (by $G$) if it is an eigenstate of \emph{all} $(N,{\mathcal H}_s)$-postselected scattering matrices that are compatible with $G$, i.e. that commute with the set of operators in $G$. The vacuum state $\ket{0}$ and all states with $\mathrm{dim}({\mathcal H}_s)=1$ are trivially symmetry-protected since postselection projects on the one-dimensional space spanned by the state itself. The notion becomes interesting, however, for $N\geq1$ and $\mathrm{dim}({\mathcal H}_s)\geq2$, which ensures that postselection projects on a subspace of dimension greater than $1$. In that case, most states are not protected.

There are two reasons why a state $\ket{\psi}$ may fail to be protected. First, photons may be scattered between the modes in ${\mathcal H}_s$, performing an ($\hat{S}$-dependent) quantum operation. Postselection (to $N$ photons in the modes ${\mathcal H}_s$) is insensitive to these changes and the postselected state is different from the input, hence not protected. This can be resolved by using a different subspace ${\mathcal H}_{s}'$ in which at least one basis mode is uniquely characterized by quantum numbers preserved by all $\hat{S}$ compatible with all the elements in $G$. Then, it is straightforward to write down $N$-photon Fock states that are protected. Since all scattering matrices commute with the symmetry operators in $G$, the corresponding quantum numbers cannot be changed by $\hat{S}$. Therefore, if a vector $\ket{\psi} = a_{\psi}^\dag\ket{0}$ in ${\mathcal H}_s'$ is uniquely defined by preserved quantum numbers, then any state $(a_\psi^\dag)^N \ket{0}$ is $({\mathcal H}'_s)$-symmetry-protected. Note that here postselection projects on a high-dimensional Hilbert space ($N$ photons in $\mathrm{dim}({\mathcal H}'_s)$ modes) and that if $\hat{S}$ were not compatible with $G$ (and if $\psi$ were not the unique mode in $\mathcal{H}_s'$ with the given preserved quantum numbers), then this state would in general not be an eigenstate of the postselected scattering matrix. These protected states are all Fock states and are all eigenstates of some symmetry operators.

In all the previous cases, one might just as well postselect on the one-dimensional initially populated subspace spanned by the protected state, since none of the other states in $\mathcal{H}_s'^N$ will be populated through scattering (by construction). However, as we will see, this type of protection can be extended to superposition states and whole subspaces in which postselection on $N$ photons in $\mathcal{H}_s$ brings a genuine advantage. In this case, a second source of decoherence has to be taken into account: the probability that photons are scattered out of the modes in $\mathcal{H}_s$ is, in general, different for different modes, which would change an initial superposition state in $\hat{S}$-dependent (and, thus, unknown) ways. Similarly, different states may acquire different phase shifts. And since both mechanisms depend on unknown details of $\hat{S}$, they will lead to decoherence.

In the following, we construct states that are protected against both sources of decoherence in scattering problems with cylindrical symmetry, where $G$ comprises the rotations around a symmetry axis and mirror reflections at a plane containing it. We construct different classes of entangled protected states and discuss some uses of the states found.

\section{Properties of the eigenmodes of angular momentum and helicity}
\label{sec:ourstates}
\begin{figure}[ht]
    \subfloat[Subfigure 1 list of figures text][]{
        \includegraphics[width=0.45\textwidth]{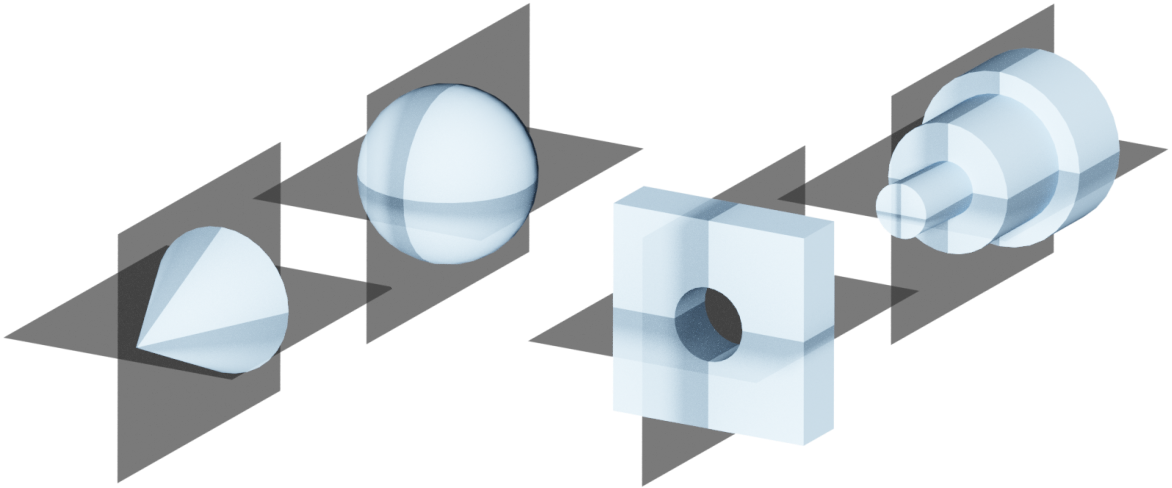}
        \label{mirrors}}
    \qquad
    \subfloat[Subfigure 2 list of figures text][]{
        \includegraphics[width=0.45\textwidth]{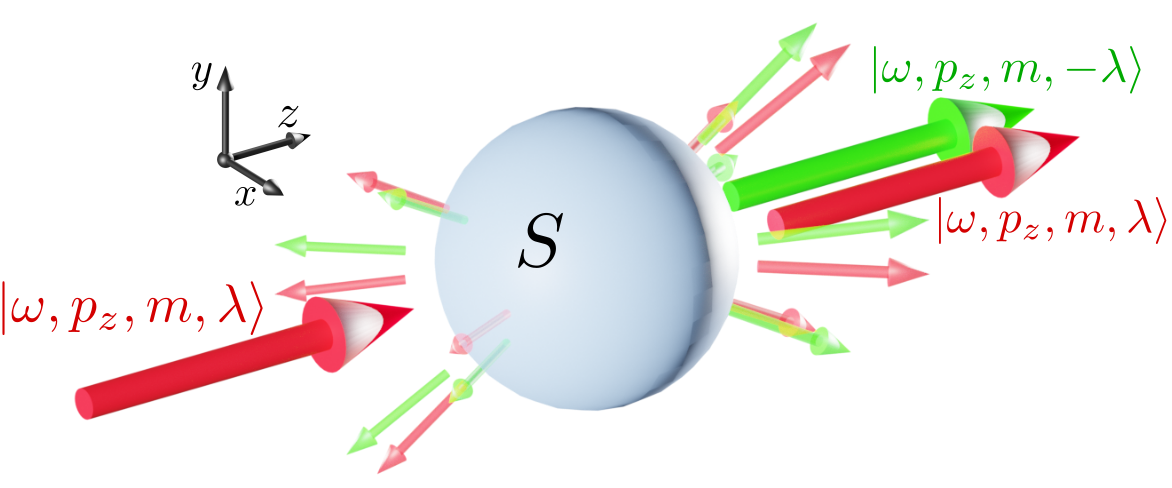}
        \label{non_dual_sphere}}
    \caption{In (a) cylindrically symmetric scatterers are depicted. In (b), the splitting of the helicity of a Bessel mode is shown when it interacts with a non-dual scatterer. Modes of same frequency, $\omega$, and linear momentum, $p_z$, are postselected at the output.}
\end{figure}

Let us consider photonic eigenstates of one component of the total angular momentum, $J_z=L_z+S_z$, and helicity ($\Lambda=\mathbf{J}\cdot\mathbf{p}/p$), where $L_z$ and $S_z$ are, respectively, the $z$ components of the orbital (OAM) and spin (SAM) angular momenta~(\cite{Messiah}, Chapter XIII), $\mathbf{p}$ is the linear momentum operator and $p$ its modulus. Now, we label the eigenstates with the eigenvalue of $J_z$, $m=\{-\infty,...,-1,0,1,...,\infty\}$, and the sign of the eigenvalue of $\Lambda$, $\lambda=\{-1,+1\}$. Therefore, our set of electromagnetic modes can be labeled as $\vec{E}_{m,\lambda}(\vec{x},t)$, where $\vec{E}$ is the electric field associated with this particular mode, and we will drop the spatio-temporal dependence of the mode from now on. As we are concerned only with the symmetries of our system, we are leaving out other degrees of freedom which would uniquely define the electromagnetic mode. In principle, one could also use the optical frequency, $\omega$ and the $z$ component of the linear momentum, $p_z$, and this would define the set of Bessel modes $\vec{E}_{\omega,p_z,m,\lambda}$ (see Fig.~\ref{non_dual_sphere})~\cite{PhysRevA.86.042103}, or the optical frequency and $j$, the quantum number of the square of the total angular momentum, $J^2$, forming the set of multipolar modes $\vec{E}_{\omega,j(j+1),m,\lambda}$~\cite{rose}. For our purposes it is sometimes convenient to use, instead of the helicity eigenstates, the eigenstates of the mirror transformation $M_y$, describing reflection at the $xz$ plane, a symmetry of the scatterers we consider; we label them with their eigenvalue $\tau=\{1,-1\}$.

In the following we consider cylindrically symmetric scatterers, that is, $G=\{M_y, R_z(\theta)=e^{i\theta J_z} : \theta\in[0,2\pi)\}$ is formed by the rotations around $z$ axis and reflections at the $xz$ plane as mentioned before. Note that, in this case, $G$ is the point group $C_{\infty v}$. For $\mathcal{H}_s$ we take the space spanned by all Bessel modes with fixed frequency $\omega$ and linear momentum, $p_z$. To construct the protected states, we look at subspaces of $\mathcal{H}_s$ which map to themselves under the action of rotations around the $z$ axis and the mirror transformation. More specifically, in this work we consider the spaces spanned by the bases
\begin{equation}
    \mathcal{H}_0 = \mathrm{span}\Big\{ \vec{E}_{0, +}\; ;\; \vec{E}_{0, -} \Big\}
    \label{H0}
\end{equation}
and
\begin{equation}
    \mathcal{H}_{m} = \mathrm{span}\Big\{ \vec{E}_{m, +}\; ;\;  \vec{E}_{m, -}\; ;\; \vec{E}_{-m, +}\; ;\; \vec{E}_{-m, -} \Big\}.
    \label{Hm_-m}
\end{equation}

\begin{figure}[hbt!]
    \centering
    \subfloat[Subfigure 1 list of figures text][]{
        \includegraphics[width=0.45\textwidth]{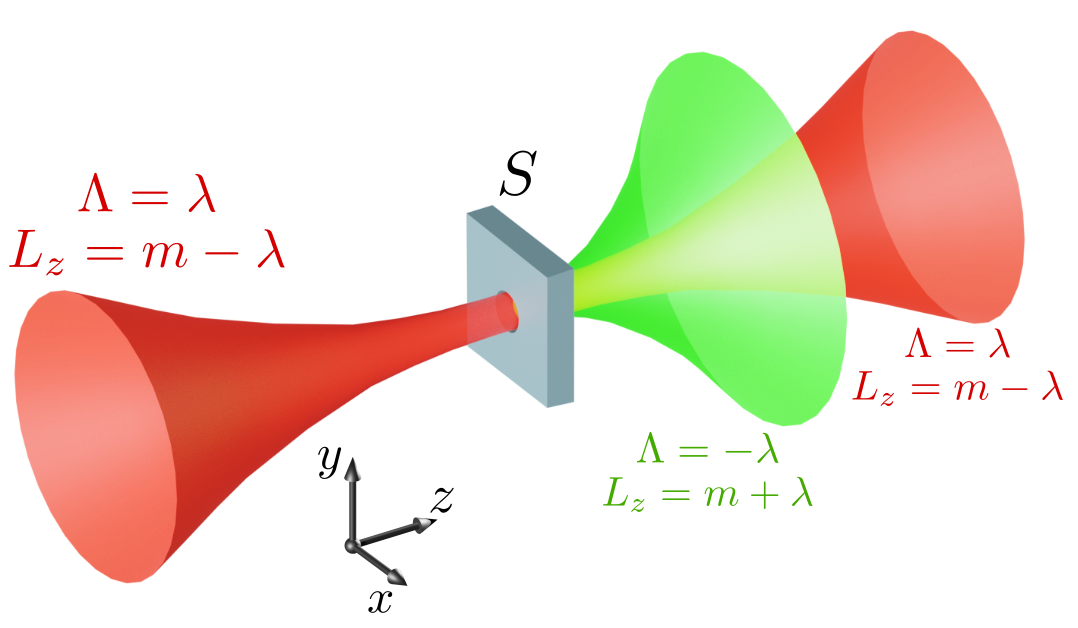}
        \label{beams_oam}}
    \qquad
    \subfloat[Subfigure 2 list of figures text][]{
        \includegraphics[width=0.45\textwidth]{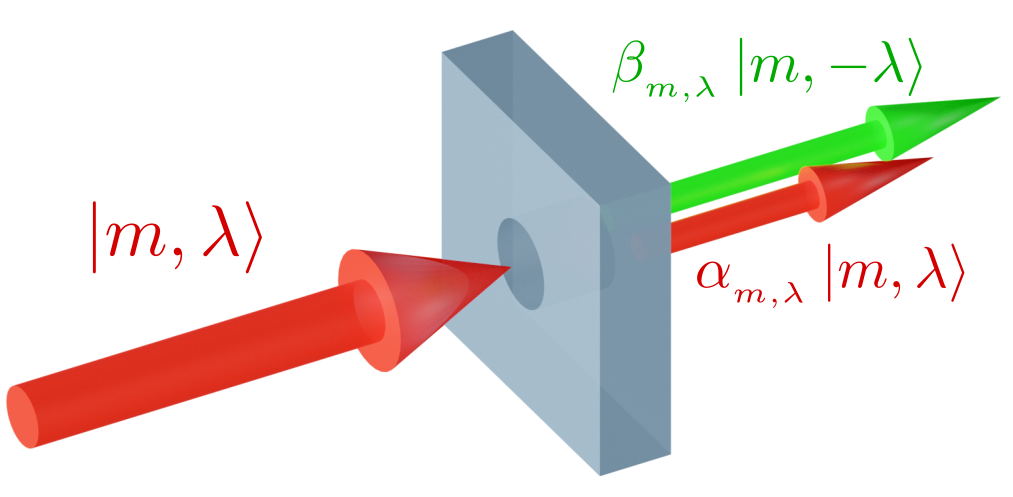}
        \label{photon_state_am}}
    \caption{Interaction of light with a cylindrically symmetric scatterer. In (a), a classical beam with OAM $m-\lambda$ and helicity $\lambda$ is focused on the scatterer. There are two output beams: one with the same components of OAM and helicity, and another one, with a difference of two units of OAM and opposite helicity. In (b), a single-photon state with angular momentum $m$ and helicity $\lambda$ interacts with the scatterer. At the output, a superposition of states with the same and opposite helicities is found, with probability amplitudes $\alpha_{m,\lambda}$ and $\beta_{m,\lambda}$, respectively.}
    \label{scatt_fig}
\end{figure}

Let us briefly remind of the form that relevant single-particle operators take in these subspaces. In the case of the Hilbert space $\mathcal{H}_0$, the $z$ component of angular momentum operator is $J_z = \text{diag}(0,0)$ and the mirror operator is
\begin{equation}
	M_y = \left(\begin{matrix} 0 && 1\\ 1 && 0 \end{matrix} \right),
	\label{MyH0}
\end{equation}
while the postselected scattering operator (or input-output relations, see Fig.~\ref{scatt_fig}) for a cylindrical target is given for this space by:
\begin{equation}\label{eq:S0}
	S = \left(\begin{matrix} \alpha && \beta\\ \beta && \alpha \end{matrix} \right),
\end{equation}
with $\alpha,\beta \in \mathbb{C}$. For the space given in Eq.~\eqref{Hm_-m} $J_z=\mathrm{diag}(m,m,-m,-m$), the mirror operator can be written as
\begin{equation}
	M_y = \left(\begin{matrix} 0 && 0 && 0 && 1\\ 0 && 0 && 1 && 0\\ 0 && 1 && 0 && 0 \\ 1 && 0 && 0 && 0 \end{matrix} \right)
	\label{MyHm}
\end{equation}
and the scattering operator is
\begin{equation}
    S = \left(\begin{matrix} \eta && \zeta && 0 && 0 \\ \epsilon && \gamma && 0 && 0 \\ 0 && 0 && \gamma && \epsilon \\ 0 && 0 && \zeta && \eta \end{matrix} \right),
    \label{S}
\end{equation}
with $\eta,\zeta,\epsilon,\gamma \in \mathbb{C}$. Note that any operator, $S$, defined in this way, fixes the whole dynamics of the scattering problem by defining the linear response of the considered input modes. This implies that the evolution of any input state (even in the multiphotonic case) is grounded in the single-photon nature of the interaction.

An important goal of this study is to find states of light which are symmetry-protected, i.e. states that are left invariant by all scattering operators which commute with $J_z$ and $M_y$ (here and in the following ``left invariant" always refers to the state after postselection). One can check at once that the single-photon eigenstates of $M_y$ in the space given by Eq. (\ref{H0}), fulfill this condition, i.e.
\begin{equation}
S \left(\vec{E}_{0,+} + \tau \vec{E}_{0,-}\right)=s_\tau \left(\vec{E}_{0,+} + \tau \vec{E}_{0,-}\right)~~(\tau=\pm1),
\label{mirroreig}
\end{equation}
where $s_\tau = \alpha + \tau\beta$. 

\section{Interaction of two-photon states with cylindrical samples}
\label{sec:2-photon-states}

We proceed by motivating the general case with the simple case of two-photon states. It was experimentally proved in \cite{buse2018symmetry} that there is one two-photon state which, when interacting with a circular nanoaperture, remains unaffected. This state is a simultaneous eigenstate of the angular momentum operator and mirror operator. For the particular case of modes with $m=0$, measured in Ref.~\cite{buse2018symmetry}, the mirror operator and the angular momentum commute. However, this is not true in the general case of modes with arbitrary angular momentum $m$. Therefore, we divide the section in two subsections: the study of modes in $\mathcal{H}_0$ and $\mathcal{H}_m$. When dealing with photon states we will use Fock state notation. In the case of $\mathcal{H}_0$ we will use $\ket{n_1,n_2}$, where $n_1$ ($n_2$) is the occupation of the mode with positive (negative) helicity, except when noted. On the other hand, when considering space $\mathcal{H}_m$, the notation will be $\ket{n_1,n_2,n_3,n_4}$. Each of the $n_i$ occupation numbers refers to the modes in $\mathcal{H}_m$ following the order expressed in Eq.~\eqref{Hm_-m}.

\subsection{Two photons in $\mathcal{H}_0$}\label{n2}

For two indistinguishable photons in the modes in $\mathcal{H}_0$, there is a three-dimensional state space given by:
\begin{equation}
	\mathcal{H}^{N = 2}_0 = \mathrm{span}\Big\{ \ket{1, 1}, \ket{2, 0}, \ket{0, 2} \Big\}.
	\label{basis_2_m0}
\end{equation}
It can be readily seen that this specific basis for $\mathcal{H}^{N = 2}_0$ is made of eigenstates of helicity, but the states do not have a well-defined mirror eigenvalue, $\tau$. Due to its importance in the scattering of cylindrically symmetric systems, let us study the properties of the mirror operator. Thus, we construct the $\hat{M}_y$ operator in $\mathcal{H}^{N = 2}_0$ from Eq. \eqref{MyH0}. Then, the transformation of Fock space vectors in Eq. \eqref{basis_2_m0} under the mirror operator is given in matrix form by:
\begin{equation}
	\hat{M}_y = \left(\begin{matrix} 1 && 0 && 0\\ 0 && 0 && 1\\ 0 && 1 && 0 \end{matrix} \right)
\end{equation}
(note that we have chosen the notation $\hat{O}$ to represent a generic Fock space operator, whereas the hatless form $O$ is reserved for the mode operators). If we diagonalize this matrix, we obtain the following set of eigenvalues and orthonormal eigenvectors which also constitute a complete basis set for $\mathcal{H}^{N = 2}_0$:
\begin{align}
	\ket{\Phi_1} &= \ket{1,1}~~(\tau = 1)\\
	\ket{\Phi_2} &= \frac{1}{\sqrt{2}}\left(\ket{2,0} + \ket{0,2}\right)~~(\tau = 1)\\
	\ket{\Phi_3} &= \frac{1}{\sqrt{2}}\left(\ket{2,0} - \ket{0,2}\right)~~(\tau = -1).
	\label{sym_prot0}
\end{align}
Two mirror symmetric and one antisymmetric states are found. The mirror antisymmetric state is uniquely characterized by conserved quantum numbers (total angular momentum and mirror eigenvalues) and, thus, it is protected under postselected scattering. On the other hand, the two mirror symmetric states, in principle, could be mixed after undergoing the scattering process (and it is easy to construct a scattering operator that does so) as they both share the $\tau = 1$ quantum number. Thus, $\ket{\Phi_3}$ is an example of an entangled two-photon state which is symmetry-protected under the scattering from an arbitrary cylindrical sample.

There is another approach which leads to the same result, but that allows us to find two other states which also are two-photon protected states. Instead of starting with eigenmodes of helicity given in Eq. \eqref{H0}, one can redefine the single-photon Hilbert space basis and use the eigenstates of the mirror operator given in Eq.~\eqref{mirroreig}. With this approach one obtains three symmetry-protected states for the two-photon case we are studying, which are:
\begin{align}
    \ket{S_1} &= \frac{1}{2}\left(\ket{2,0} + \sqrt{2}\ket{1,1} + \ket{0,2}\right)\\
    \ket{S_2} &= \frac{1}{2}\left(\ket{2,0} - \sqrt{2}\ket{1,1} + \ket{0,2}\right) \label{twop}
\end{align}
and the previously obtained $\ket{\Phi_3}$ state. Interestingly, one finds that all three of them are Fock states in the protected modes given in Eq.~\eqref{mirroreig} ($\ket{2,0}',\ket{0,2}'$, and $\ket{1,1}'$, respectively, where the sign "$~'~ $" is used to specify that the mirror eigenbasis is being used, see Section~\ref{n4}). Let us remark here that this is a general consequence of the single-particle nature of the scattering, i.e. that if $a_k^\dag\ket{0}$ are protected then so are $\Pi_k (a_k^\dag)^{m_k}\ket{0}$. For brevity, we sometimes refer to the latter state as a ``product of the states $a_k^\dag\ket{0}$''. In conclusion, $\ket{S_1}$, $\ket{S_2}$ and $\ket{\Phi_3}$, are symmetry-protected because they can be written as products of protected single-photon states. This is a particularity of the $\mathcal{H}_0$ space that will be more deeply analyzed in the next section. As we will show later, every protected state with $N$ photons in the modes which span $\mathcal{H}_0$ can be written in the same fashion.

\subsection{Two photons in $\mathcal{H}_m$}
When the modes under consideration have $m\neq 0$, the situation is a bit more complex, due to the fact that the mirror operator does not commute with the angular momentum operator on $\mathcal{H}_{m\not=0}$. As before, we start with the space given by the modes in Eq.~(\ref{Hm_-m}). The necessity of including states of negative angular momentum is now obvious as we want to consider a subspace that the mirror operator leaves invariant. In this case, the accessible part of Fock space is ten-dimensional:
\begin{equation}
\begin{split}
	\mathcal{H}^{N = 2}_m = \mathrm{span}\Big\{
	& \ket{2,0,0,0}, \ket{1,1,0,0}, \ket{1,0,1,0}, \ket{1,0,0,1},\\
	& \ket{0,2,0,0}, \ket{0,1,1,0}, \ket{0,1,0,1}, \ket{0,0,2,0},\\
	& \ket{0,0,1,1}, \ket{0,0,0,2} \Big\}.
\end{split}
\label{H2m}
\end{equation}
The elements in Eq.~\eqref{H2m} can be separated in subspaces with different $m_{tot}$. This can be done because, in a basis of angular momentum eigenmodes, the eigenvalues of the second quantized total angular momentum of the field are $m_{tot} = \sum\nolimits_i m_i$, which give the set of values: $0,~2m,~-2m$. The elements of each of these subspaces are, respectively:
\begin{align}
	\nonumber
	\mathcal{S}_0 &= \mathrm{span}\Big\{ \ket{1,0,0,1}, \ket{0,1,1,0}, \ket{1,0,1,0}, \ket{0,1,0,1} \Big\}\\
	\nonumber
	\mathcal{S}_{+} &= \mathrm{span}\Big\{ \ket{2,0,0,0}, \ket{1,1,0,0}, \ket{0,2,0,0} \Big\}\\
	\nonumber
	\mathcal{S}_{-} &= \mathrm{span}\Big\{ \ket{0,0,2,0}, \ket{0,0,1,1}, \ket{0,0,0,2} \Big\}.
\end{align}
It can be noted that the only subspace which is invariant (whose elements transform to other elements of the subspace) under the action of the mirror operator is $\mathcal{S}_0$. Therefore, states belonging to subspace $\mathcal{S}_0$ are the only ones which can have simultaneously well-defined angular momentum and mirror eigenvalues. Now, transformations under the mirror operator are given by Eq.~\eqref{MyHm}, which allows us to construct the mirror operator matrix for the $\mathcal{S}_0$ subspace as:
\begin{equation}
	\hat{M}_y = \left(\begin{matrix} 1 && 0 && 0 && 0\\ 0 && 1 && 0 && 0\\ 0 && 0 && 0 && 1\\ 0 && 0 && 1 && 0 \end{matrix} \right).
\end{equation}
whose eigenvectors and eigenvalues are:
\begin{align}
	\ket{\Psi_1} &= \ket{1,0,0,1}~~(\tau = 1) \label{sym_unp_1}\\
	\ket{\Psi_2} &= \ket{0,1,1,0}~~(\tau = 1) \label{sym_unp_2}\\
	\ket{\Psi_3} &= \frac{1}{\sqrt{2}}\left( \ket{1,0,1,0} + \ket{0,1,0,1} \right)~~(\tau = 1) \label{sym_unp_3}\\
	\ket{\Psi_4} &= \frac{1}{\sqrt{2}}\left( \ket{1,0,1,0} - \ket{0,1,0,1} \right)~~(\tau = -1). \label{sym_prot}
\end{align}
\begin{figure}[ht]
    \centering
    \subfloat[Subfigure 2 list of figures text][]{
        \includegraphics[width=0.45\textwidth]{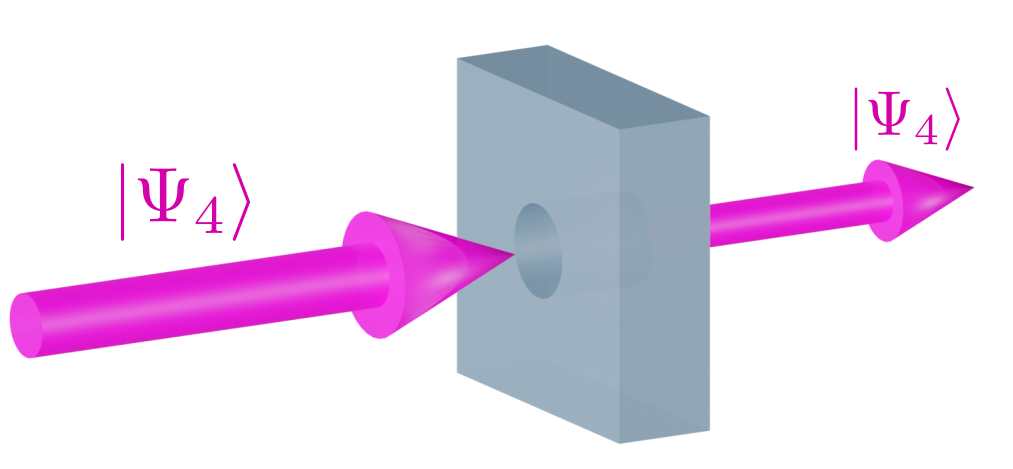}
        \label{sym_prot_im}}
    \qquad
    \subfloat[Subfigure 1 list of figures text][]{
        \includegraphics[width=0.45\textwidth]{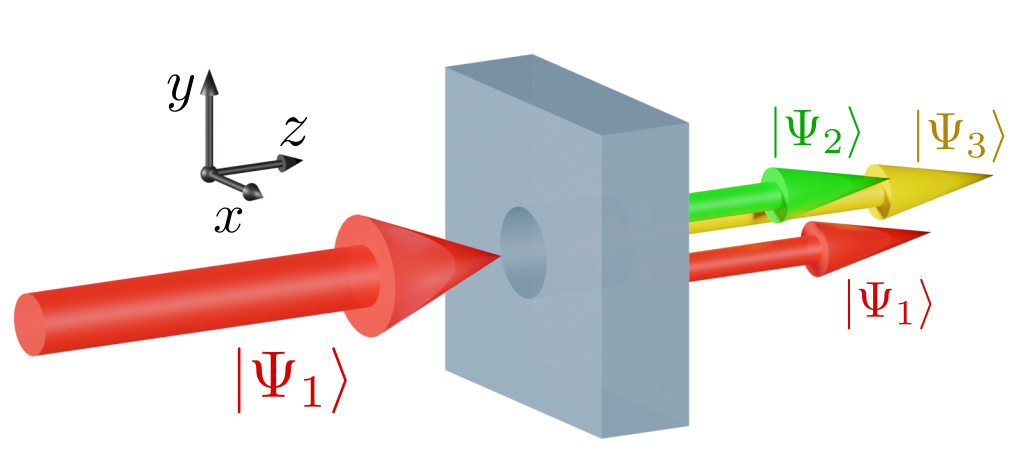}
        \label{sym_unp_im}}
    \caption{Schematic behavior of the states in Eqs.~\eqref{sym_unp_1}-\eqref{sym_prot} interacting with a cylindrically symmetric scatterer. In (a), the only mirror antisymmetric state, $\ket{\Psi_4}$, is left invariant through scattering. In (b), a mirror symmetric eigenstate, $\ket{\Psi_1}$, generates through scattering a superposition of all the mirror symmetric states.}
\end{figure}

As in the case of $\mathcal{H}_0$, the state $\ket{\Psi_4}$ will not mix (under postselected scattering) with other states, either belonging to spaces with a different $m^\prime$ or the other three mirror symmetric states in the same $\mathcal{S}_0$ subspace. Therefore, for every $m$, the mirror antisymmetric states generated in this way are protected and do not mix with any other by scattering on a cylindrically symmetric sample (Fig.~\ref{sym_prot_im}). Also, it is easy to check that none of the mirror symmetric states that diagonalize the scattering matrix are independent of the scattering coefficients, in other words, symmetry arguments alone cannot warrant their protection (Fig.~\ref{sym_unp_im}).

Finally, notice that in the single-photon Hilbert space $\mathcal{H}_m$, symmetry-protected states cannot be found. Notwithstanding, in the two-photon case such states exist. This is a consequence of quantum interference and, thus, it is a feature of the multiphotonic nature of the states we are considering.

\subsection{Summary of two-photon interactions with cylindrically symmetric objects}

All light modes can be classified according to their angular momentum and helicity. This classification block-diagonalizes the scattering matrix of cylindrically symmetric objects in submatrices given by modes in spaces $\mathcal{H}_0$ and $\mathcal{H}_m$. We have seen that for pairs of photons we can always find subspaces $\mathcal{S}_0$ where the total angular momentum of the state is zero. Importantly, each of these subspaces contains a mirror antisymmetric state which is symmetry-protected. In other words, when scattering these photon pairs off a cylindrical scatterer and postselecting for two photons, we always find the same pair: the scatterer cannot redistribute the two photons in the subspace due to conservation laws and the single-particle nature of the scattering we are considering (see \ref{appendA}).

It may be interesting to point out, that the two-photon protected states we have identified are entangled according to standard criteria for entanglement of indistinguishable particles used in the literature. State $\ket{\Phi_3}$ of Eq.~(\ref{sym_prot0}), for instance, has a Slater number $2$ and, thus, it can be considered particle-entangled according to \cite{ESBL02}. Nevertheless, it can be written as a product state between the mirror-symmetric and anti-symmetric in $\mathcal{H}_0$ and, therefore, it is not entangled according to most definitions \cite{Silberhorn,LZLL01}. In contrast, state $\ket{\Psi_4}$ of Eq.~(\ref{sym_prot}) cannot be written as a single product of creation operators applied to the vacuum state in any way (neither with orthogonal nor with non-orthogonal modes) and, thus, it is entangled according to all these definitions \cite{Silberhorn,ESBL02,LZLL01}.

Finally, while for two-photon states this procedure has been quite direct, there are still a few questions which remain open. The obvious one is: can we generalize this procedure to arbitrary multiphoton states? In the next section we proceed to generalize our study of symmetry-protection to $N$-photon states.

\section{Interaction of multiphoton states with cylindrical samples}\label{sec:N-photon-states}

The search for symmetry-protected states in the multiphoton case is, in general, much more complicated. As the number of particles increases, all the eigenspaces of interest in which to search for protected states increase in dimension, making it harder to find or exclude solutions. In  particular, the simple sufficient condition for protection --being a state uniquely characterized (within the postselected space) by $\hat{J}_z$ and $\hat{M}_y$ eigenvalues loses its usefulness as all the simultaneous eigenspaces become degenerate for $N>2$.

One can, however, dig into the formal definition of symmetry-protection and try to make it operative. A mathematical procedure to construct or exclude $N$-photon symmetry-protected states is presented in \ref{appendB} based on this idea. While the basic reasoning can be used for any type of scattering problem under symmetry constraints, here we exploit features of the cylindrical symmetry that impose specific relations between the eigenvectors and eigenvalues of cylindrically symmetric scattering matrices. We use it to prove that there are no other symmetry-protected states in $\mathcal{H}^N_m$ apart from products of the state given by Eq.~\eqref{sym_prot}.

In what follows, we proceed as before, by studying symmetry-protection separately for $\mathcal{H}_0^N$ and $\mathcal{H}_m^N$ spaces.

\subsection{$N$ photons in $\mathcal{H}_0$} \label{n4}

As explained earlier, to understand symmetry-protection in $\mathcal{H}^N_0$, we should begin with the set of single-photon modes which are joint eigenstates of $M_y$ and $J_z$ operators, i.e.,
\begin{equation}
	\ket{1, 0}' = \frac{1}{\sqrt{2}}\Big( \ket{1, 0}  + \ket{0, 1} \Big)
	\label{m0_newbasis}
\end{equation}
and
\begin{equation}
	\ket{0, 1}' = \frac{1}{\sqrt{2}}\Big( \ket{1, 0}  - \ket{0, 1} \Big).
	\label{m0_newbasis1}
\end{equation}
Eq.~\eqref{mirroreig} shows that states in Eqs.~\eqref{m0_newbasis}-\eqref{m0_newbasis1} are left invariant when impinging on a cylindrical sample. Thus, one can construct multiphoton states which are protected by defining creation and annihilation operators for these states and taking their products as pointed out in Section~\ref{sec:general}. Denoting by $\hat{a}^{\dag}_{0, s}\ket{0}$, the mirror symmetric state in Eq. \eqref{m0_newbasis}, and $\hat{a}^{\dag}_{0, a}\ket{0}$, the mirror antisymmetric mode in Eq. \eqref{m0_newbasis1}, we can identify symmetry-protected states of $N$ photons in the following way:
\begin{equation}
	\ket{n_{s}, n_{a}}' = \prod_{t=s,a}\frac{\left( \hat{a}^{\dag}_{0, t} \right)^{n_{t}}}{\sqrt{n_{t}!}}\ket{0},
	\label{sym_protH0N}
\end{equation}
where $N=n_a+n_s$ and $n_{s}$ ($n_{a}$) is the occupation number of the mirror symmetric (antisymmetric) photon mode. All these states have well-defined angular momentum and mirror transformations. In particular, their mirror eigenvalue is given by $(-1)^{n_{a}}$.

Finally, just for completeness, when $N$ is odd there are $(N+1)/2$ mirror symmetric and $(N+1)/2$ mirror antisymmetric states of this kind. However, in the case of $N$ being even, there are $N/2$ mirror antisymmetric states, and $N/2+1$ symmetric states. In both cases, the total number of states is $N + 1$.

\subsection{$N$ photons in $\mathcal{H}_m$}\label{n3}

Following the reasoning of the previous sections, we know that products of symmetry-protected states are also protected. Therefore, a state of the form
\begin{equation}
\ket{\Psi} = \left(\hat{a}_{m,+}^\dag \hat{a}_{-m,+}^\dag - \hat{a}_{m,-}^\dag \hat{a}_{-m,-}^\dag\right)^{N/2}\ket{0}
\label{prot_HmN}
\end{equation}
must be left invariant by any cylindrically symmetric scatterer. Note that this state belongs to the $\mathcal{S}_0$ subspace of the $N$-photon Fock space and its mirror symmetry depends on whether $N/2$ is even or odd. In general, products of such states constructed from different $m$ and $N$ values are also protected, even the products of these states and the ones obtained in Eq. \eqref{sym_protH0N}.

Interestingly, the state given in Eq. \eqref{prot_HmN} is the only symmetry-protected state that can be obtained for a fixed value of $m$ and $N$. This can be proved from the very general definition of symmetry-protection given in Section \ref{sec:general}, exploiting the properties of the eigenstates of cylindrically symmetric scattering matrices. The details are given in \ref{appendB}. The proof rests on the defining property that a protected state is required to be an eigenstate of \emph{all} scattering operators $S, S',\dots$, etc. compatible with the group of symmetry operators $G=\{M_y, R_z(\theta)=e^{i\theta J_z} : \theta\in[0,2\pi)\}$. Importantly, this constraint not only determines the possible form of any compatible scattering the matrix, as given by Eq. \eqref{S}, but also the transformations between the eigenmodes of two compatible scattering matrices. Finally, we observe that the transformations between two infinitesimally distinct scattering matrices $S$ and $S'$ suffice to prove that the symmetry-protected state in Eq. \eqref{prot_HmN} is unique.

\section{Symmetry-protection and decoherence-free subspaces}\label{sec:DFS}

Note that so far we have discussed the protection of \emph{one-dimensional} subspaces, namely single multiphoton states that are preserved under scattering when postselecting on subspaces of ${\cal H}_0^N$ and ${\cal H}_m^N$ $N$-photon Fock spaces with null total angular momentum. While this provides an interesting characterization of the scatterer and may be useful for certain applications, it is not sufficient to transmit qubits or other forms of quantum information, for which at least a two-dimensional protected subspace is required. But there is no way that cylindrical symmetry alone can guarantee that after postselection a state like $(a P_m + b P_{m'})\ket{0}$ is unchanged, where $P_m\ket{0}$ and $P_{m'}\ket{0}$ represent $N\geq1$-photon protected states as constructed above. Symmetry arguments alone cannot warrant that the scattering transformation of the states is independent of $m$: while the use of protected states ensures that the transformation is proportional to the identity, both the amplitude and the phase may depend on $m$, and thus both the relative phase and amplitude of $a$ and $b$ can change, decohering the qubit.

However, as we now show, with one additional assumption on the scatterer, decoherence-free subspaces may be constructed. Moreover, we show that the construction of these subspaces is possible even in the case where losses are present. In what follows we consider that this is, in fact, the case.

If the scatterer (and hence the scattering matrix) can be considered to be static, i.e. constant during a time interval $[t_1,t_2]$, then a protected state $P_m\ket{0}$ scattered at time $t_1$ or time $t_2$ will undergo exactly the same transformation (loss in amplitude and phase change) and, therefore, if we can postselect on no losses, any superposition of the two would be unaffected. Indeed, for the protected states constructed in the previous sections, the state after scattering (but before postselection) is of the form $\lambda P_m\ket{0} + \ket{\psi_R}$, where $\lambda$ is the eigenvalue of the protected state and $\ket{\psi_R}$ is the part in which at least one photon has been scattered into environmental modes. If the scattering is time-independent and Markovian, sending an input state in a superposition of being in the first or the second time-bin $(a P_m(t_1) + b P_m(t_2))\ket{0}$ will be scattered into $\lambda (a P_m(t_1) + b P_m(t_2))\ket{0} + a \ket{\psi_R(t_1)} + b\ket{\psi_R(t_2)}$. Postselecting on having the input number of photons, $N$, either in the first and zero in the second time-bin or vice versa will yield the unchanged input state. In principle, this postselection can be done without affecting the superposition, e.g., by filtering on the correct photon number ($N$) in the full set of employed modes and an integer multiple of $N$ of photons in each time-bin (the point of this latter measurement is to ensure that the $N$ photons all appear in a single time-bin without learning in which one). Thus, the whole \emph{two-dimensional subspace} is transmitted in protected fashion.

Furthermore, one can generalize this to construct a \emph{$d$-dimensional decoherence-free subspace} given by: $\sum_{i=1}^d a_i P_m(t_i)\ket{0}$, as long as the scattering matrix remains static in the time interval $[t_1, t_d]$. Since the loss of probability only depends on the total photon number, not on the number of time-bins, these do not suffer larger losses (but require more demanding postselection). Note that the simplest realization is the use of single-photon ($N=1$) protected states as given in Eq.~(\ref{mirroreig}), in which case the qubit is just a suitable angular-momentum choice of the time-bin qubit long used in quantum communications \cite{BGTZ99} and for which efficient quantum logic has been developed (e.g., \cite{Roh14}).

The protection we consider here is, of course, not protecting against
photon losses, but it is a postselected protection: we identify a $d\geq2$-dimensional subspace of $N$-photon states within which \emph{all} states are transmitted with fidelity 1 provided that $N$ photons have been transmitted. Consequently, one can view the scattering process as the action of a quantum erasure channel \cite{GBP97}: either the transmitted state is lost (if postselection fails) or the state is transmitted perfectly. These channels are known to have a finite quantum capacity of $1-2\varepsilon$, where $\varepsilon$ is the loss probability for losses below 50\% \cite{BDS97} and can, therefore, be used to transmit quantum information or distribute entanglement~\cite{valivarthi2016,wengerowski2019}. If two-way classical communication between sender and receiver is possible, the quantum capacity is increased to $1-\varepsilon$, i.e. it is larger than zero except for 100\% losses. Erasure errors allow for more efficient quantum error correction that can tolerate large loss rates \cite{GBP97,Kni03,MZ+Liang17}. 

\section{Conclusions}\label{sec:conclusion}

In conclusion, we have set the general stage in which the notion of symmetry-protection can be analyzed and better understood. We have specialized on the case of angular momentum states of light which are left invariant in scattering problems with cylindrical symmetry. In addition, we have shown that protected states emerge, not only in the one or two-photon level, but also in the general multiphoton case. Finally, we have proposed the superposition of time-bin symmetry-protected states as a suitable candidate to generate \emph{$d$-dimensional} decoherence-free subspaces for quantum communication applications.

\appendix

\section{Symmetry-protection and the single-particle nature of scattering} \label{appendA}
Along this study we have focused our analysis on scattering processes involving a passive linear scatterer, whose action can be fully understood on the single-particle space, spanned by the relevant modes of the photons. For instance, if Bessel modes are considered, we fix $\omega$ and $p_z$ (by initial preparation and postselection) and only consider the angular momentum and helicity quantum numbers $m$ and $\lambda$.

In general, a passive and linear scatterer is described by a unitary matrix $T$ on the space of all modes; postselecting to the modes of fixed $\omega,~p_z$ (e.g., identical to the initial ones) selects the sub-block $(T)_{\omega p_z,\omega p_z}$ which is only constrained to have rows and columns of norm $\leq 1$, but can otherwise be arbitrary; let's denote it by $S$ in the following. The emergence of $S$ can also be understood in terms of an underlying (microscopic) Hamiltonian which is, after all, the generator of the time evolution of the system. However, our approach is not microscopic. Instead, input-output relations are considered, defined as transition probabilities between states prepared in the distant past and states detected in the distant future (with respect to the moment in which the interaction actually takes place).

Components of the $S$ matrix depend, in general, on the specific details of the interaction between the scatterer and the optical modes. However, some relations can be found among them when considering a physical sample which has some symmetries. Choosing cylindrical symmetry, for instance, implies that $S$ must be invariant under rotations around an axis, in our case the $z$ axis. Consequently, $S$ must be block-diagonal in the angular momentum basis, i.e. $S=\oplus_m S_m$, where $S_m$ acts on the subspace of all modes with angular momentum $m$. In the case we consider, these blocks are two-dimensional $S_m=(S_m)_{\lambda,\lambda'}$ with $\lambda,\lambda'=\pm$. Moreover, in most situations, a cylindrical object has also associated a mirror symmetry. The way in which this symmetry is reflected in the components of $S$ is: $(S_m)_{\lambda,\lambda'}=(S_{-m})_{-\lambda,-\lambda'}$. It is this property that allows us to identify multiphoton states that are left invariant under \emph{all} cylindrically symmetric scatterers.

The transformation that creation operators undergo when interacting with a passive linear scatterer (after postselection) is given by 
\begin{equation}
\hat{a}^{\dag}_{m,\lambda} \mapsto \sum_{\nu=\pm} (S_m)_{\nu,\lambda}\hat{a}^{\dag}_{m,\nu},
\end{equation}
and if $N$-photon states are considered, 

\begin{align}
\nonumber
\hat{a}^{\dag}_{m_1,\lambda_1}\hat{a}^{\dag}_{m_2,\lambda_2}...\hat{a}^{\dag}_{m_N,\lambda_N} \mapsto \sum_{\nu_1,\nu_2,...,\nu_N} &(S_{m_1})_{\nu_1,\lambda_1}(S_{m_2})_{\nu_2,\lambda_2}...(S_{m_N})_{\nu_N, \lambda_N}\hat{a}^{\dag}_{m_1,\nu_1}\hat{a}^{\dag}_{m_2,\nu_2}...\hat{a}^{\dag}_{m_N,\nu_N}.
\end{align}
Concretely, for the two-photon case and choosing $m' = -m$ one gets the following transformation
\begin{equation}
\hat{a}^{\dag}_{m,\lambda}\hat{a}^{\dag}_{-m,\lambda'} \mapsto \sum_{\nu,\nu'} (S_m)_{\nu,\lambda}(S_{-m})_{\nu',\lambda'}~\hat{a}^{\dag}_{m,\nu}\hat{a}^{\dag}_{-m,\nu'}.
\end{equation}
More specifically, the transformation for the state in Eq. \eqref{sym_prot} is given by
\begin{align}
\ket{\Psi_4} \mapsto \sum_{\nu,\nu'} &\Big\{(S_m)_{\nu,+}(S_{-m})_{\nu',+} - (S_m)_{\nu,-}(S_{-m})_{\nu',-}\Big\}\hat{a}^{\dag}_{m,\nu}\hat{a}^{\dag}_{-m,\nu'}\ket{0},
\label{psi4_append}
\end{align}
which can be further simplified if one considers the symmetry condition $(S_m)_{\lambda,\lambda'}=(S_{-m})_{-\lambda,-\lambda'}$ which ensures that the terms with $\nu \neq \nu'$ cancel. Taking this into account, the transformation in Eq. \eqref{psi4_append} can be written as
\begin{align}
    \ket{\Psi_4} \mapsto \sum_{\nu = \pm} C_\nu ~\hat{a}^{\dag}_{m,\nu}\hat{a}^{\dag}_{-m,\nu}\ket{0},
\end{align}
with $C_\nu = (S_m)_{\nu,+}(S_{m})_{-\nu,-} - (S_m)_{\nu,-}(S_{m})_{-\nu,+}$. It is easy to check that $C_- = -C_+ \equiv -C$, giving as a result
\begin{equation}
    \ket{\Psi_4} \mapsto  C ~(\hat{a}^{\dag}_{m,+}\hat{a}^{\dag}_{-m,+}-\hat{a}^{\dag}_{m,-}\hat{a}^{\dag}_{-m,-})\ket{0},
\end{equation}
which shows that symmetry-protection of state in Eq. \eqref{sym_prot}
can be understood based on the single-particle nature of the interaction and the relations between scattering coefficients imposed by the symmetry of the problem. Finally, the symmetry-protection arising in states given in Eqs.~\eqref{m0_newbasis} and \eqref{m0_newbasis1}, can directly be understood from the single-particle nature of the interaction, as they are themselves single-particle symmetry-protected states.


\section{Proof on the uniqueness of symmetry-protection in $\mathcal{H}^N_m$} \label{appendB}

In the space of modes $\mathcal{H}_m$, a postselected single-particle scattering matrix $S=S_{m}\oplus S_{-m}$ given by Eq.~(6) is block-diagonal and the $2\times2$ matrix $S_{m}$ has two eigenvalues $\nu_{m,\pm}$ and corresponding eigenmodes $v_{m,\pm}$. In general, they are not orthogonal to each other, but for a generic scattering matrix $S$ they are linearly independent and the two eigenvalues are distinct. Due to mirror symmetry, each eigenvalue $\nu_{m,\pm}$ is twofold degenerate with one eigenvector belonging to the $m$ subspace and the second one belonging to $-m$. Since the scattering matrix $S$ has block-diagonal form, the degenerate eigenvectors on the two subspaces are related by a flip operation $X$, $S_{-m}= X S_m X$. Since $m$ is fixed for the remainder of this Appendix, we simply write $\nu_{\pm}$.

The $N$-photon Fock space $\mathcal{H}^N_m$ can be separated in $N+1$ degenerate eigenspaces of the scattering operator $\hat{S}$ with eigenvalues $\nu_{+}^N$, $\nu_{+}^{N-1}\nu_-, \dots, \nu_{+}\nu_{-}^{N-1}, \nu_{-}^{N}$. In principle, each eigenspace could contain symmetry-protected states, but as we show below all the states outside the $(\nu_+\nu_-)^{N/2}$-subspace depend on the details of $S$ and, hence, none of them can be symmetry-protected. Moreover, we show that the protected states can without loss of generality be chosen as $\hat{J}_z$ eigenstates. Finally, we demonstrate that within the simultaneous $\hat{J}_z$ and $\hat{S}=(\nu_{+}\nu_{-})^{N/2}$ eigenspace, there is a unique symmetry-protected $N$-photon state, i.e., the one given in Eq.~\eqref{prot_HmN} of the main text (with $J_z=0$ and simultaneously an $M_y$ eigenstate).

By definition, a symmetry-protected state must be an eigenstate of \emph{all} cylindrically symmetric scattering operators. To see a given eigenspace of $\hat{S}$ does not contain protected vectors, it suffices to show that for every vector $\ket{\Psi}$ in that space, there is another cylindrically symmetric scattering operator, $\hat{S}'$, so that $\ket{\Psi}$ does \emph{not} lie in any of its eigenspaces. For the case at hand, this can be seen for any $S'$ with eigenvectors distinct from those of $S$. Then, we can express (without loss of generality) the eigenmodes of $S$ through those of $S'$ as $v_{m,\nu_\pm}=p_{\pm}v_{m,\nu'_\pm}+q_{\pm}v_{m,\nu'_\mp}$, where $p_{\pm},q_{\pm}\not=0$ (as we can take $S'$ to have two linearly independent eigenvectors distinct of those of $S$, i.e., $v_{m,\nu_\pm} \not = v_{m,\nu'_+},v_{m,\nu'_-}$) and $\nu'_\pm$ refers to the two possible eigenvalues of $S'$ operator. When denoting the creation operators associated to these modes we will use the notation $\hat{a}^\dag_{\pm,\pm}$ and $\hat{a}'^\dag_{\pm,\pm}$. The first subscript refers to the sign of the angular momentum $m$ and the second to the scattering eigenvalue $\nu_\pm$ or $\nu'_\pm$, respectively.

In general, for the $N$-photon eigenspace of the $\hat{S}$ scattering operator belonging to the eigenvalue $\nu_+^M\nu_-^{N-M}$, any state can be written as
\begin{equation}
  \label{eq:phiNM}
  \ket{\Psi_{M,N-M}} = \sum_{l_+=0}^M\sum_{l_-=0}^{N-M} x_{l_+,l_-}
  (\hat{a}_{+,+}^\dag)^{M-l_+}   (\hat{a}_{-,+}^\dag)^{l_+}
  (\hat{a}_{+,-}^\dag)^{N-M-l_-}   (\hat{a}_{-,-}^\dag)^{l_-}\ket{0},  
\end{equation}
where $x_{l_+,l_-}$ is a coefficient that depends on $l_\pm$. A first simplification is that since $\hat{S}$ and $\hat{J}_z$ commute we can choose $\ket{\Psi_{M,N-M}}$ always as an eigenstate of $\hat{J}_z$. (If a protected state consists of a superposition of $J_z$-eigenstates to different eigenvalues, then each eigencomponent must itself be protected, since $S$ does not couple or mix the components.) Each term in the sum of Eq. \eqref{eq:phiNM} is a $\hat{J}_z$ eigenstate with eigenvalue $\hat{J}_z/|m|= N-2(l_++l_-)$, thus, for $\ket{\Psi_{M,N-M}}$ to be an $\hat{J}_z$-eigenstate, $K = l_+ + l_-$ must be a constant. For a joint $\hat{S}$ and $\hat{J}_z$ eigenstate we then write
\begin{align}
  \ket{\Psi_{M,N-M,K}} &= \sum_{l_+=0}^M\sum_{l_-=0}^{N-M} \delta_{l_+ + l_- = K} ~ x_{l_+,l_-}
  (\hat{a}_{+,+}^\dag)^{M-l_+}   (\hat{a}_{-,+}^\dag)^{l_+}
  (\hat{a}_{+,-}^\dag)^{N-M-l_-}   (\hat{a}_{-,-}^\dag)^{l_-}\ket{0},
\end{align}
and we distinguish the cases (a) $M\leq N-M$ and (b) $M> N-M$ which lead to slightly different expressions for $\ket{\Psi_{M,N-M,K}}$ related to our arbitrary choice of ordering the $\nu_\pm$ eigenvalues. We discuss case (a) in the following, case (b) reduces to (a) when changing the labelling of the eigenvalues $\nu_+\leftrightarrow\nu_-$.

In case (a), $l_+$ can run over all available values while satisfying the $K =
l_++l_-$ constraint, and we have:
\begin{align}
  \label{eq:11a}
  \ket{\Psi_{M,N-M,K}} &= \sum_{l=0}^{\min\{M,K\}} x_{l,K-l}
  (\hat{a}_{+,+}^\dag)^{M-l}   (\hat{a}_{-,+}^\dag)^{l}
  (\hat{a}_{+,-}^\dag)^{N-M-K+l}   (\hat{a}_{-,-}^\dag)^{K-l}\ket{0}.
\end{align}
This state can be rewritten in terms of the modes associated with another generic symmetric scattering operator, $\hat{a}'^\dag_{\pm,\pm}$, using the transformation relations between the $S$ and $S'$ eigenmodes. Pulling out the $p_\pm$ factor and defining for convenience $\tilde{q}_\pm=(q_\pm/p_\pm)$, $x'_{l,K-l} = x_{l,K-l}p_+^{M}p_-^{N-M}$ and $K_M=\min\left\{ K,M \right\}$, the state given by Eq.~\eqref{eq:11a} can be written as
\begin{align}\label{eq:phiprime}
\nonumber  \ket{\Psi_{M,N-M,K}} &= \sum_{l=0}^{K_M} x'_{l,K-l}
\sum_{r_1=0}^{M-l}\sum_{r_2=0}^{l}\sum_{r_3=0}^{N-M-K+l}\sum_{r_4=0}^{K-l} {M-l\choose r_1}{l\choose r_2}{N-M-K+l\choose r_3}{K-l\choose r_4}\tilde{q}_+^{r_1+r_2}\tilde{q}_-^{r_3+r_4}\\ 
&\hspace*{3ex}(\hat{a}'^\dag_{+,+})^{M-l-r_1+r_3}(\hat{a}'^\dag_{-,+})^{l-r_2+r_4}(\hat{a}'^\dag_{+,-})^{N-M-K+l-r_3+r_1}(\hat{a}'^\dag_{-,-})^{K-l-r_4+r_2}\ket{0}
\end{align}
in the $S'$ eigenbasis. The scattering eigenvalues to which the summands with index $\vec{r}=(r_1,r_2,r_3,r_4)$ belong is $(\nu'_+)^{M-g}(\nu'_-)^{N-M+g}$, where $g=r_1+r_2-(r_3+r_4)$. Let us take a look to the terms with $\vec{r}=0$. They give a vector in the
$(\nu'_+)^M(\nu'_-)^{N-M}$ eigenspace. Given that we have chosen that $p_\pm$ is non-zero, the $\vec{r}=0$ only vanishes if $x_{l,K-l} = 0$, i.e., provided that the initial vector $\ket{\Psi_{M,N-M,K}} = 0$. That is to say, $\ket{\Psi_{M,N-M,K}}$ necessarily has a non-zero component in the $(\nu'_+)^M(\nu'_-)^{N-M}$ eigenspace. Since for it to be symmetry-protected it must be an $S'$ eigenvector, we can conclude that all components outside of the $(\nu'_+)^M(\nu'_-)^{N-M}$ eigenspace must vanish. This constrains the  $x_{l,K-l}$ a protected state can have: they must be chosen such that all terms outside that eigenspace vanish. This must hold independently of $p_{\pm},q_{\pm}$ since $S$ and $S'$ are arbitrary and thus all pairs $(\tilde{q}_+,\tilde{q}_-)\in\mathbbm{C}^2$ can occur. In other words, the terms in the state given by Eq.~\eqref{eq:phiprime} which are proportional to different powers of $\tilde{q}_\pm$ and belong to $S'$-eigenspaces characterized by $g\not=0$ must individually vanish.

An especially simple case to consider (and sufficient for our proof) is the one of infinitesimally different $S$ and $S'$ scattering operators: in that case, we have that $\tilde{q}_\pm$ correspond to infinitesimal displacements and we can focus only on the first-order terms, neglecting the higher-order ones. In the state in Eq. \eqref{eq:phiprime}, there are two terms proportional to $\tilde{q}_+$ and another two for $\tilde{q}_-$. The former are:
\begin{align*}
 &= \sum_{l=0}^{K_M}
   x'_{l,K-l}(M-l) \ket{F_{M-l-1,l,N-M-K+l+1,K-l}}\\
  &{}+\sum_{l=0}^{K_M} x'_{l,K-l} l \ket{F_{M-l,l-1,N-M-K+l,K-l+1}},
\end{align*}
where we have used the notation 
\begin{equation}
    \ket{F_{k,l,m,n}}=(\hat{a}'^\dag_{+,+})^k(\hat{a}'^\dag_{-,+})^l(\hat{a}'^\dag_{+,-})^m(\hat{a}'^\dag_{-,-})^n\ket{0}. 
\end{equation}
Notice that two vectors $\ket{F_{k,l,m,n}}$ and $\ket{F_{k',l',m',n'}}$ are linearly independent  unless all indices are the same. These vectors are, in general, neither orthogonal nor normalized, but that is of no importance below. One can check that the Fock state corresponding to the $l=K_M$ term in the first sum does not appear in the second, while all the others do. On the other hand, the $l=0$ term in the second sum vanishes. We can then relabel the terms in the second sum (by replacing $l\to l'+1$ and summing over $l'=0,\dots,K_M-1$) to get
\begin{align*}
  &\sum_{l=0}^{K_M-1} x'_{l+1,K-l-1} (l+1) \ket{F_{M-l-1,l,N-M-K+l+1,K-l}}
\end{align*}
so that the two sums can be put together:
\begin{align*}
  &(M-K_M)x'_{K_M,K-K_M}\ket{F_{M-K_M-1,K_M,N-M-K+K_M+1,K-K_M}}\\ 
  &{}+\sum_{l=0}^{K_M-1} \left[x'_{l,K-l}(M-l) + x'_{l+1,K-l-1} (l+1)
     \right] \ket{F_{M-l-1,l,N-M-K+l+1,K-l}}.
\end{align*}
For this term to be zero, all summands have to vanish separately. That implies
\begin{align}
  \label{eq:12}
  \nonumber
  (M-K_M)x'_{K_M,K-K_M}&=0,\\
  \nonumber
  (M-K_M+1)x'_{K_M-1,K-K_M+1}&={}-(K_M)x'_{K_M,K-K_M},\\
  \nonumber
  (M-K_M+2)x'_{K_M-2,K-K_M+2}&={}-(K_M-1)x'_{K_M-1,K-K_M+1},\\
  \nonumber
  \dots&\dots\\
  (M)x'_{0,K}&=-(1)x'_{1,K-1}.
\end{align}
Thus, unless $K_M = M$ the coefficient $x'_{K_M,K-K_M}=0$ and, subsequently, all other $x'_{l,K-l}=0$. Before considering the case $K_M=M$, let's turn to the first-order terms in $\tilde{q}_-$. It can be treated similarly and yields
\begin{align*}
  &(N-M-K)x'_{0,K} \ket{F_{M+1,0,N-M-K-1,K}}\\
  &{}+\sum_{l=1}^{K_M}(N-M-K+l)x'_{l,K-l}
    \ket{F_{M-l+1,l,N-M-K+l-1,K-l}}\\
  &\hspace*{-0mm}+(K-K_M)x'_{K_M,K-K_M}\ket{F_{M-K_M,K_M+1,N-M-K+K_M,K-K_M-1}} \\
    &{}+\sum_{l=0}^{K_M-1} (K-l)x'_{l,K-l}\ket{F_{M-l,l+1,N-M-K+l,K-l-1}}
\end{align*}
Now, there are two unpaired terms and the $l'$ term of the second sum matches the $l'+1$ term of the first one, which leads to the set of conditions
\begin{align}
  \label{eq:13}
  \nonumber
  (K-K_M)x'_{K_M,K-K_M}&=0,\\
  \nonumber
  (N-M-K)x'_{0,K} &=0,\\
  \nonumber
  (N-M-K+1)x'_{1,K-1} &= -(K)x'_{0,K},\\
  \nonumber
  (N-M-K+2)x'_{2,K-2} &= -(K-1)x'_{1,K-1},\\
  \dots&\dots\nonumber\\
  (N-M-K+K_M)x'_{K_M,K-K_M}&=-(K-K_M+1)x'_{K_M-1,K-K_M+1}.
\end{align}
Note that now, unless $N-M-K=0$, the second of these conditions fixes $x'_{0,K}=0$ and then all other $x'_{l,K-l}=0$, including $x'_{K_M,K-K_M}$.

In summary, unless $K - K_M = 0$ and $N - M - K = 0$ there cannot exist a symmetry-protected state. Recall that we work in the case (a) $M < N-M$, hence if $K = N-M$, then, $\min(M,K) = M$, $K = M = N-M$ and, consequently, $K = M = N/2$. This means that all possible symmetry-protected states, $\ket{\Psi_{M,M,M}}$, fulfill the relations $\hat{J}_z\ket{\Psi_{M,M,M}}=0$ and $\hat{S}\ket{\Psi_{M,M,M}} = (\nu_+\nu_-)^{N/2}\ket{\Psi_{M,M,M}}$. In other words, all symmetry-protected states belong both to the subspace of full Fock space spanned by states with null angular momentum and the $(\nu_+\nu_-)^{N/2}$ scattering eigenspace.

It remains to consider the case $K=M=N/2$ (note that these conditions only allow for an even number, $N$, of photons) and to prove that the $N$-photon protected pairs given by Eq.~\eqref{prot_HmN} saturate the whole set of possible symmetry-protected states $\ket{\Psi_{M,M,M}}$. This is straightforward: one quickly confirms that for $M = K = K_M = N/2$, Eq.~\eqref{eq:12} and Eq.~\eqref{eq:13} give the same set of conditions, namely
\begin{equation}
  \label{eq:14}
  lx'_{l,K-l} =-(K - l + 1) x'_{l-1, K-l+1}\,\,(l=1,\dots,K),
\end{equation}
which yields
\begin{equation}
  \label{eq:15}
  x'_{l, K-l} = (-1)^l{K\choose l}x'_{0,K}.
\end{equation}
Recall that the $x'_{l,K-l}$ differ from the unprimed coefficients only by a constant factor $(p_+p_-)^K$. So far, all these are given in the eigenbasis of an arbitrary symmetric scattering matrix and hold in \emph{any} such basis. Thus, for simplicity, we finally choose $S=\text{diag}(1,-1,-1,1)$ in the helicity basis which is one of the admissible generic scattering matrices. Then the two $\nu_+$ modes are $a_{+m,+}$ and $a_{-m,-}$ and the two $\nu_-$ modes are $a_{+m,-}$ and $a_{-m,+}$. Plugging the expression obtained for the $x_{l,K-l}$ coefficients into Eq.~\eqref{eq:11a}, we observe that the subset of states $\ket{\Psi_{M,M,M}}$, omitting the common and constant factor $x_{0,K}$, can be written as:
\begin{align}
  \label{eq:17}
  \ket{\Psi_{M,M,M}} &= \sum_{l=0}^K (-1)^l {K \choose l}
  (\hat{a}_{+,+}^\dag)^{K-l}   (\hat{a}_{-,+}^\dag)^{l}
  (\hat{a}_{+,-}^\dag)^{l}   (\hat{a}_{-,-}^\dag)^{K-l}\ket{0}\nonumber\\
  &= \left[ \hat{a}_{+,+}^\dag \hat{a}_{-,-}^\dag  - \hat{a}_{-,+}^\dag \hat{a}_{+,-}^\dag\right]^{N/2}\ket{0} \nonumber\\
  &=\left[ \hat{a}_{+m,+}^\dag \hat{a}_{-m,+}^\dag  - \hat{a}_{m,-}^\dag
    \hat{a}_{-m,-}^\dag\right]^{N/2}\ket{0},
\end{align}
i.e. exactly the protected pair identified in Eq. \eqref{prot_HmN} of the manuscript. \qed


\begin{thebibliography}{99}
\bibitem{helicity} Helicity is a magnitude which permits the generalization of the concept of polarization to the regime in which the propagation of light is non-paraxial. It is the conserved quantity associated with the duality symmetry of electromagnetic fields.
\end{thebibliography}
\end{document}